\documentclass[twocolumn]{article}
\usepackage{graphicx,amsmath,amssymb,longtable,url}
\usepackage[latin9]{inputenc}
\newcommand{\be}{\begin{equation}}
\newcommand{\ee}{\end{equation}}
\newcommand{\bea}{\begin{eqnarray}}
\newcommand{\eea}{\end{eqnarray}}

\renewcommand{\a}{{\bf a}}
\newcommand{\Areg}{{\cal A}}
\newcommand{\ahat}{\hat{\bf a}}

\renewcommand{\b}{{\bf b}}
\newcommand{\Breg}{{\cal B}}
\newcommand{\bhat}{\hat{\bf b}}

\newcommand{\bth}{{\bf \Theta}}

\newcommand{\A}{{\cal A}}

\newcommand{\argmax}{\mbox{argmax}}

\newcommand{\opone}{\leavevmode\hbox{\small1\kern-3.3pt\normalsize1}}
\newcommand{\para}[1]{{\vspace{0.5ex} \noindent \bf #1.}}

\begin{document}

\title{Bayesian analysis of biological networks: clusters, motifs, cross-species correlations}

\author{Johannes Berg and Michael L\"assig \\
Institut f\"ur Theoretische Physik,
Universit\"at zu K\"oln, 
Z\"ulpicherstr. 77, 50937 K\"oln, Germany\\
}
\maketitle


The complexity of an organism is only weakly linked with its number of
genes. \textit{Homo sapiens} has about 25000 genes and the roundworm \textit{C. elegans} 
about 19000~\cite{Stein:2004,Claverie:2004}, despite the different
level of complexity.  Not only the gene numbers
are similar, the genes themselves are frequently shared
across species.  Even distantly related organisms have a high fraction
of genes which stem from their common ancestor (orthologs): more than
$90\%$ of genes are shared between human and mouse and at
least $30\%$ of genes of the yeast \textit{S. cerevisiae} have orthologs
in human~\cite{eugenes_homology}.

This result is an important outcome of the recent 
genome sequencing projects. It has put the spotlight on the \textit{interactions} 
between genes: changes in the complex networks of gene
regulation, or in the interactions between proteins, may be a major
cause of phenotypic variation, more so than changes in the genes
themselves~\cite{King:1975}. The molecular basis of these interactions
includes specific binding sites on
regulatory DNA and binding domains in proteins. Binding sites can change
quickly generating new interactions or deleting old
ones~\cite{Tautz:2000,Wray:2003,BergWillmannLaessig:2004,Gelfand:2006}.

The resulting interest in biological interactions has been matched
by the development of novel experimental techniques to measure 
protein--DNA interactions
and protein--protein interactions. In particular, 
\textit{high-throughput} methods have been developed, facilitating measurements on a
genome-wide scale rather than for individual genes. Some of
the ingenious methods 
for experimentally determining biological interactions will 
be briefly reviewed in the next section. 

This experimental development is akin to the transition from sequencing small parts of
the DNA of an organism to the determination of full genomes. The growth of 
sequencing capabilities has been driving the 
development of computational methods for sequence analysis for the
past three decades. Virtually all methods for sequence analysis rely
on statistics as a tool to infer function. Examples are the detection
of genes, or of regulatory modules, or the identification of correlations
between evolutionarily related sequences~\cite{durbin.etal:1998}.

The corresponding development of computational network
biology is still in its infancy. It will require new tools
to address specific issues of biological networks.  These
are characterized by a peculiar interplay of stochasticity
and function, and in many ways epitomize our current lack
of understanding of biological systems. With  this caveat,
the point of view we take in this article is that
statistics will again  play a decisive role in our
understanding of network biology, and we point out some
currently available links  between network statistics and function.
The merit of a statistical approach may not seem obvious 
from an engineering perspective, where networks are seen as
deterministic processing machines producing a well-defined
input--output  relation. Indeed, biological networks
sometimes work in a surprisingly deterministic way: for
example, a network of a few dozen major genes generates a
well-defined spatiotemporal development pattern in the 
eukaryotic embryo. However, the underlying network
structures are fundamentally stochastic, since they arise
from the manifold tinkering and feedback processes of
biological evolution. Explaining deterministic function
from a stochastic evolution requires a statistical, dynamical
theory. 

One important aspect of this challenge is to predict
different functional units in networks. 
Different functions are reflected in a different
evolutionary dynamics, and hence in different statistical
characteristics of network parts. In this sense, the {\em global
statistics} of a biological network, e.g., its connectivity 
distribution, provides a background, and {\em local
deviations} from this background signal functional units.
In the computational analysis of biological networks, we thus  
typically have to discriminate between different statistical
models governing different parts of the dataset. The nature 
of these models depends on the biological question asked. 
We illustrate this rationale here with three examples:
identification of functional parts as highly connected
\textit{network clusters}, finding \textit{network motifs},
which occur in a similar form at different places in the
network, and the analysis of 
\textit{cross-species network correlations}, which reflect 
evolutionary dynamics between species. 

\begin{figure}[tbh!]
\includegraphics*[width=1. \linewidth]{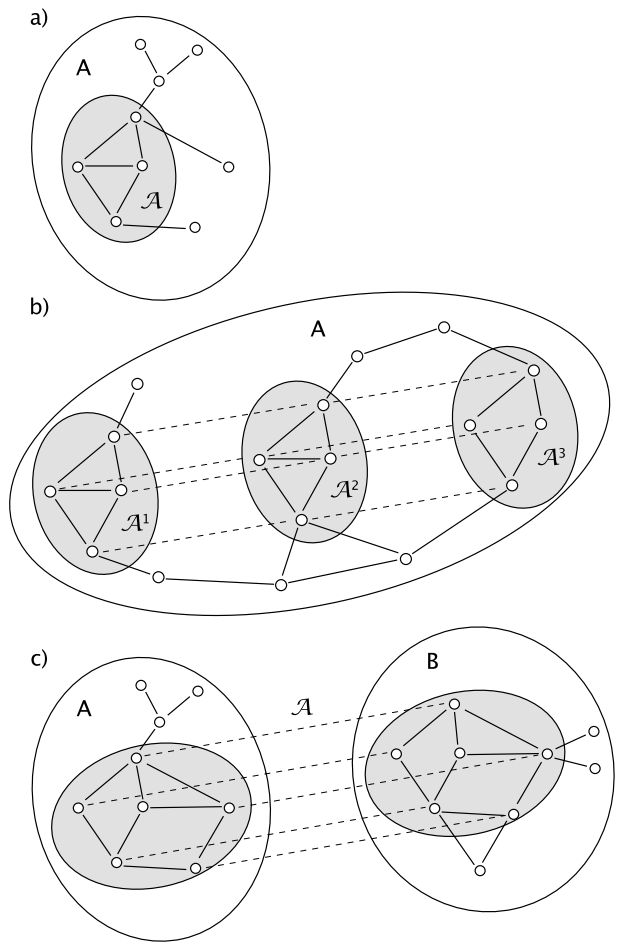}
\caption{\label{fig_stats} \small
{\bf Deviation from a uniform global statistics in biological networks.} 
a) A network cluster is distinguished by an enhanced number of intra-cluster 
interactions. 
b) A network motif is a set of subgraphs with correlated interactions. 
In a limiting case, all subgraphs have the same topology. 
c) Cross-species correlations characterize evolutionarily conserved parts of 
networks.   
}
\end{figure}

\subsection*{1. Measuring biological networks}

A wide range of experimental methods has 
been developed to measure interactions between proteins, interactions between 
proteins and regulatory DNA, and expression levels of genes. Only a brief 
review is possible here. 

In the yeast-two-hybrid (Y2H) method, the pairwise interaction between two
proteins is tested by creating two fusion
proteins~\cite{Uetz.etal:2000}. One protein is constructed with a
DNA-binding domain attached to its end, and its potential binding
partner is fused to an activation domain. If the two proteins
interact, the binding will form a transcriptional activator (generally
consisting of a DNA-binding domain and an activation domain). The
presence of an intact activator leads to the transcription of an
easily detectable reporter gene. (The reporter gene may for instance produce 
a fluorescent protein.) In principle, the amount of the
reporter gene produced can serve as a measure of the affinity between
the two proteins. The Y2H-method has been used to measure the 
protein interaction networks of 
yeast~\cite{Uetz.etal:2000}, 
\textit{C. elegans}~\cite{Li.etal:2004},
\textit{D. melanogaster}~\cite{Giot.etal:2003}, and 
human~\cite{Rualetal:2005}. The Y2H-datasets are known to 
contain a large number of false positive and false negative results.  
False negatives arise when the fusion proteins fail to localize in 
the yeast nucleus, or fail to fold properly once the new domains are attached.
False positives may be linked to high expression levels of the 
hybrid in yeast, which are never reached \textit{in vivo}. 
 
Alternative approaches include pull-down assays, where one protein
type is immobilized on a gel, and `pulls down' binding partners from a
solution. Binding partners may then be identified by various tags.
Mass spectrometry is also used to identify the interacting protein
pairs identified by such an affinity
analysis~\cite{YingmingZhao:1996}. While more accurate than the
Y2H-method, these approaches have not yet been scaled up to provide
high throughputs.

Binding of proteins, specifically transcription factors, to regulatory
DNA has long been investigated by electrophoresis, where the motility
of a DNA-fragment is altered by a protein bound to it. Chromatin
immunoprecipitation (ChIP) is an alternative procedure, which uses
specific antibodies to isolate a protein and then amplifies DNA that 
may have been isolated together (co-precipitated) with the protein. By 
running many such experiments in parallel on a microarray, this method 
can be scaled up to high 
throughputs~(ChIP-on-chip,~\cite{HorakSnyder:2002}). 

Gene expression levels can be measured on DNA-microarrays, densely
packed samples of known nucleotides, each a few tens of base-pairs
long. Currently more than $10^6$ such samples, or probes, can be
placed on a single microarray. The array is then washed with a
fluorescently labeled sample. Binding of DNA in the sample to
complementary DNA on the probe can be detected under a microscope from
the resulting fluorescence pattern. Genome-wide expression levels can
thus be measured on a single array. Many other applications of microarrays are being
developed --- for instance microarrays to measures interactions
between transcription factors and regulatory DNA. DNA-microarrays are
also making major inroads as diagnostic tools, from characterizing the
microbial communities in dentistry~\cite{Smootetal:2005} to the
early detection of cancer~\cite{Stremmeletal:2002}.

\subsection*{2. Random networks in biology}


Randomly generated networks are very useful to analyze simple
characteristics of biological networks. For instance, typical
distances on a randomly generated network generally scale
logarithmically with the number of network nodes.  Finding such 
short distances 
also in biological network data is therefore not a surprising
result and does not require a biological explanation.
Another frequent observation in biological networks is a
distribution of node connectivities with a broad tail, which
is shared by specific ensembles of random networks. This has
motivated a number of 
statistical models explaining the connectivity distribution in terms of
the underlying evolutionary
dynamics~\cite{BarabasiAlbert:1999,Vazquezetal:2003,BergLaessigWagner:2004}. 
Thus, ensembles of random networks can
be tuned to fit certain characteristics of biological
network data. Does that mean the actual network {\em
is} random? This is clearly not the case: other observables
may differ from what is expected in the random network
ensemble, and we will see that these deviations from the
``null model'' are particularly interesting as signals of
biological function. Hence, random network
ensembles play an important role in quantifying the most
unbiased background statistics of a ``functionless''
network. Their choice is a subtle issue: it has to be
motivated by what we consider as not important for the 
biological function in question. Let us now turn to a few such models.

A network is specified by its adjacency matrix $\a =(a_{ii'})$. 
For binary networks $a_{ii'}=1$ if there is a link between 
nodes $i$ and $i'$, and $a_{ii'}=0$ if there is no link. 
Networks with undirected links are represented by a symmetric 
adjacency matrix. 
The {\em in} and {\em out connectivities} of a node,
$k^{+}_{i} = \sum_{i'} a_{i'i}$ and $k^{-}_{i} = \sum_{i'} a_{i i'}$,
are defined as the number of in- and outgoing links, respectively.
The total number of directed links is given by $K = \sum_{i,i'} a_{ii'}$. 

To focus on a specific part of the network we define an ordered
subset $\Areg$ of $n$ nodes $\{r_1,\dots r_n \}$ (see
Fig.~\ref{fig_stats}a).  The subset $\Areg$ induces a \textit{pattern}
$\ahat(\Areg)$ on the network, represented by the restricted adjacency
matrix containing only links internal to node subset $\Areg$.  $\ahat$ is
thus an $n \times n$ matrix with entries $\hat{a}_{ij} = a_{r_i r_j}$
($i,j = 1, \dots, n$).  Together, subset of nodes $\Areg$ and its pattern
$\ahat(\Areg)$ form a \textit{subgraph}.

The simplest ensemble of random networks is generated by connecting
all pairs of nodes independently with the same probability $w$. 
Given a subset of nodes $\Areg$, the probability of generating pattern 
$\ahat$ is then given by 
$
P_0(\a)= \prod_{i,i' \in \Areg}^n
                (1-w)^{1-a_{ii'}}  w^{a_{ii'}} 
$
(for undirected networks the sum is restricted to $i \leq i'$). 
This well-known ensemble, named after the
pioneers of graph theory P. Erd\H{o}s and A. R\'enyi, leads to a Poissonian
distribution of connectivities.
The only free parameter of the Erd\H{o}s--R\'enyi (ER) model, the link probability $w$
between a given pair of nodes, can be tuned so that
typical graphs taken from the ER-ensemble have the same number of links
as the empirical data. If the subset of nodes $\Areg$ contains all
$n=N$ nodes of the network, $w=K/N^2$.  
Considering connected subgraphs with $n<N$, $w$ will in general be higher than $K/N^2$. Then 
the value of $w$ can be determined by generating all connected
subgraphs of size $n$ from the empirical dataset and choosing $w$ such
that the average number of links in the ER model equals the
average number of links in connected subgraphs in the data.

However, in biological networks the connectivity distribution often
differs markedly from that of the Erd\H{o}s--R\'enyi-model.
If we have reasons to assume that a biological
function is not tightly linked to connectivity at the level 
of individual nodes, we should include the connectivity
distribution in our null model. Indeed, we can easily
construct a random ensemble matching the connectivity
distribution of the dataset. 
In this ensemble, the probability $w_{ii'}$ of finding a link between a pair
of nodes $i$, $i'$ depends on the connectivities of the nodes. 
Assuming links between different node pairs to be uncorrelated, a
given subset of nodes $\Areg$ has a pattern $\ahat$ with probability 
\be
\label{sigma_enhanced}
P_0 (\ahat) = \prod_{i,i' \in \Areg}^n
                (1-w_{ii'})^{1-a_{ii'}}  w_{ii'}^{a_{ii'}}.
\ee 
For $n=N$, when $\Areg$ includes the entire network, the
probability of finding a directed link between nodes $i$ and $i'$ is
approximately $w_{ii'}=k^{-}_{r_i} k^+_{r_{i'}}/K$, that of an
undirected link $w_{ii'}=k_{r_i}
k_{r_{i'}}/K$~\cite{Itzkovitz.etal:2003}. If we furthermore
impose the constraint that the null model describe the
statistics of a {\em connected} dataset,  the probabilities in~(\ref{sigma_enhanced}) 
are increased by a factor that can be determined from the data 
as described above. The null model constructed in this way
is maximally unbiased with respect to all patterns in the
dataset beyond its connectivity distribution.

\subsection*{3. Network clusters}

A first trace of functionality in biological networks are
strong inhomogeneities in their link statistics, which are
not captured by the null model.   Examples are protein
aggregates of several proteins held together by mutual
interactions, which show up as highly connected clusters in
protein interaction networks,  
and sets
co-regulated genes (for instance by an
oncogene~\cite{Einavetal:2005}),  leading to clusters in
co-expression networks. How can we identify these clusters
statistically? 

Clusters are subgraphs with a significantly increased number of
internal links compared to the background of the network, see
Figure~\ref{fig_stats}a). The feature distinguishing clusters is the number of
internal links, 
\be 
L(\ahat) = \sum_{i,i' \in \Areg}^n \hat{a}_{ii'} \ .
\label{L}
\ee
The statistics of clusters is then described by an ensemble
\be
\label{Qcluster}
Q_{\sigma} (\ahat) = Z_{\sigma}^{-1} \exp [\sigma L(\ahat)] \, P_0 (\ahat)  
\ee
of the same form as~(\ref{sigma_enhanced}), but with a bias towards 
a high number of internal links. The average number 
of internal links is determined by 
the value of the link reward $\sigma$. We have introduced the
normalization factor 
$Z_{\sigma} = \prod_{ii'}^n \sum_{\hat{a}_{ii'}=0,1} \exp[\sigma L(\ahat) ] \, P_0(\ahat)$, 
which ensures that $Q_{\sigma}(\ahat)$ summed over all patterns 
$\ahat$ gives unity. 

Is a given pattern $\ahat$ more likely part of a
cluster as described by the model~(\ref{Qcluster}), or is it more likely 
part of the background described by the 
null model~(\ref{sigma_enhanced})? To address this question, 
we define the so-called {\em log-likelihood score}
\be
\label{bayes_score}
S(\Areg,\sigma)=\log \left( \frac{Q_{\sigma}(\ahat)}{P_0(\ahat)} \right) = 
\sigma  L(\ahat(\Areg)) - 
\log Z_{\sigma}
\ .  
\ee 
A positive score  
results if it is more likely for the pattern $\ahat(\Areg)$ to arise in 
the model describing clusters than in the alternative null model. High scores 
indicate strong deviations from the null model. Of course this 
an attractive property for the algorithmic search for
deviations from the null model. 
As shown in the appendix, the form of the score (\ref{bayes_score}) is related in a simple way 
to the probability that pattern $\ahat$ comes from the model describing 
clusters. 

Patterns with a high score (\ref{bayes_score}) are \textit{bona fide}
clusters.  The first term of the score weighs the total number of
links. As expected, a pattern with many internal links yields a high
score. The second term acts as a threshold and assigns a negative
score to a pattern with a too small number of internal links. This term
takes into account the connectivities of the nodes: 
highly connected nodes have more internal links already in the
null model. Node subsets with highly connected nodes tend to give
lower scores. The score (\ref{bayes_score}) thus goes beyond simple
measures of clustering, such as the number of internal links,   
and provides a statistical basis for cluster
detection.

\begin{figure}[tbh!]
a)
\includegraphics*[width=.7 \linewidth]{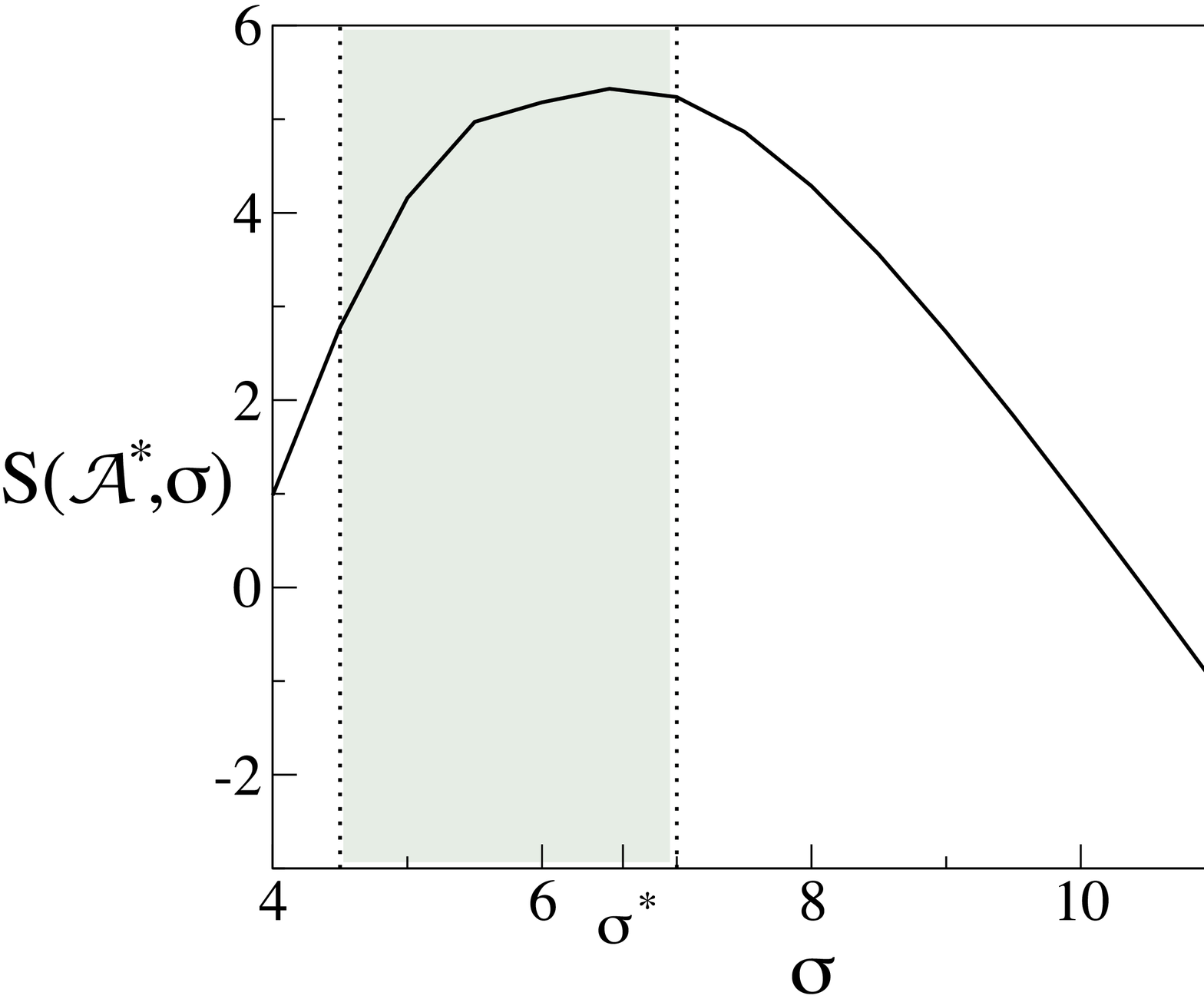}

\includegraphics*[width=1 \linewidth]{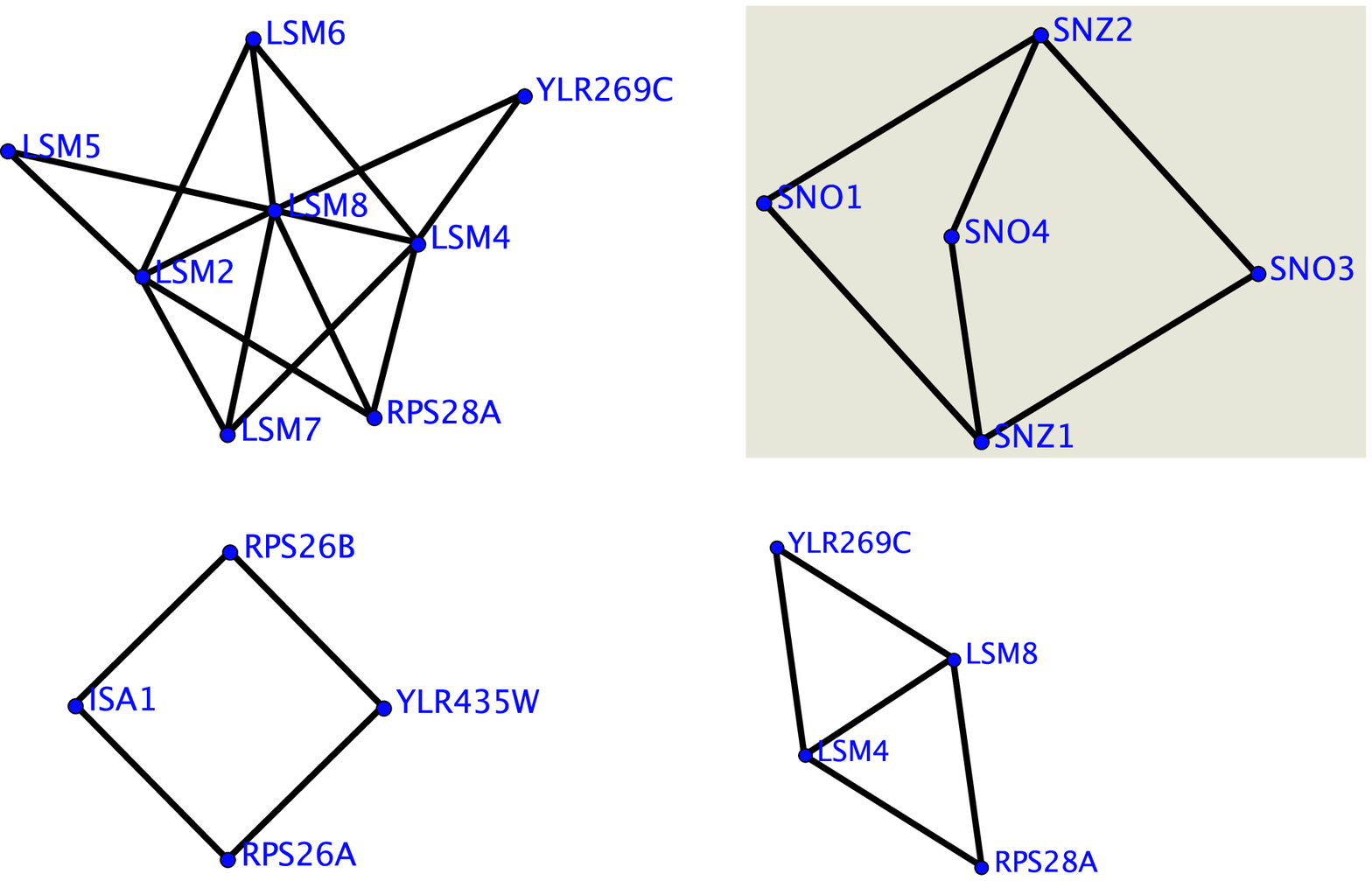}
b)
\caption{\label{fig_protclust_score_sigma} \small
{\bf Scoring clusters in protein interaction networks.} 
a) 
The score $S$ of the maximum-score node subset $\Areg^\star(\sigma)$ is 
shown as a function of the scoring parameter $\sigma$. The dotted lines 
indicate the values of $\sigma$ where the maximum-score node subset changes. 
The maximum of the score with respect to $\sigma$ indicates 
the optimal scoring parameter $\sigma^\star=6.6$. The grey region $4.25<\sigma<7$
indicates the values where $\Areg^\star(\sigma)=
\Areg^\star(\sigma^\star)$. 
b) The maximum-score subgraphs for $\sigma<4.25$, $4.25<\sigma<7$, 
$7<\sigma<11$,$\sigma>11$ (left to right). The subgraph resulting from the 
optimal scoring parameter is highlighted in grey. The maximum-score subgraphs 
for $7<\sigma<11$ and for $\sigma>11$ are distinguished by the connectivities 
of their nodes with the latter having a higher average connectivity. This 
accounts for the former having a higher score for $7<\sigma<11$ despite the 
smaller number of internal links.}
\end{figure}

Given the scoring parameter $\sigma$, the maximum-score node subset 
$\Areg^\star(\sigma)$ is defined by 
\be \Areg^\star(\sigma)=
\argmax_{\Areg}S(\Areg,\sigma) \ .  
\ee 
At this point, the scoring
parameter $\sigma$ is a free parameter, whose value needs to be
inferred from the data. This can be done by applying the principle of
maximum likelihood: $\sigma$ is determined by the requirement that the
model describing clusters~(\ref{Qcluster}) optimally describes the
statistics of the maximum-score pattern. For a given pattern $\ahat$,
the optimal fit is defined by the so-called \textit{maximum likelihood value} 
$\sigma^\star=\argmax_{\sigma} Q_{\sigma}(\ahat(\Areg))$, 
which maximizes the likelihood of generating
pattern $\ahat(\Areg)$ under the model~(\ref{Qcluster}).  Since
$\log(x)$ is a monotonously increasing function, the maximum
likelihood value $\sigma^\star$ coincides with the maximum of the
log-likelihood score (\ref{bayes_score}) over $\sigma$. The
maximum-score node subset at the optimal scoring parameter is then
determined by the joint maximum of the score over $\Areg$ and $\sigma$
\be 
S(\Areg^\star,\sigma^\star)= \max_{\sigma}
S(\Areg^\star(\sigma),\sigma) =\max_{\Areg,\sigma} S(\Areg,\sigma) \ .
\ee 
One can easily show that the maximum-likelihood value of $\sigma$
sets the expected number of links in the ensemble
$Q_{\sigma^\star}$ equal to the actual number 
of links in pattern $\ahat^\star$: setting the derivative of
(\ref{bayes_score}) with respect to $\sigma$ equal to zero gives 
\be
 \langle L(\ahat) \rangle_{Q_{\sigma^\star}} 
= L(\ahat^\star) \ . 
\ee

\para{Clusters in protein interaction networks}
We use the scoring function~(\ref{bayes_score}) to identify clusters
in the protein interaction network of yeast, namely the
high-throughput dataset of Uetz et al.~\cite{Uetz.etal:2000}. At a
given value of the scoring parameter $\sigma$, the maximum-score
node subset $\Areg^\star(\sigma)$ is identified using a simple Monte-Carlo
algorithm.  
At different values of $\sigma$, different node subsets
$\Areg^\star(\sigma)$ yield the highest score (compared to all other
node subsets). The resulting subgraphs are shown in Fig. \ref{fig_protclust_score_sigma}a). 
At low values of $\sigma$,
subgraphs with many nodes, but comparatively few internal interactions
per node yield the highest score. At high values of $\sigma$,
subgraphs with many internal interactions are favored. However these
subgraphs tend to be small. The interplay between subgraph size and
internal connectivity leads to a joint score maximum over $\Areg$ and $\sigma$ at the optimal
scoring parameter $\sigma^\star=6.6$, see
Fig. \ref{fig_protclust_score_sigma}a). 

The maximum-score cluster $\Areg^\star \equiv \Areg^\star(\sigma^\star)$ 
consists of the proteins SNZ1,SNZ2,SNO1,SNO3, and
SNO4, highlighted in grey in Fig. \ref{fig_protclust_score_sigma} b). 
The proteins in this cluster have a common function; they are
involved in the metabolism of pyridoxine and in the synthesis
of thiamin~\cite{Ensembl:2005,GO:2000}.
Furthermore, SNZ1 and SNO1 have been found to be co-regulated and their 
mRNA levels increase in response to starvation for aminoacids 
A,  
Ura, 
and Trp~\cite{Padillaetal:1998}. 

\subsection*{4. Network motifs}

The topology of a subgraph may be associated with a specific function. 
A possible example is a feed-forward loop acting as a high-frequency filter
in a regulatory network~\cite{Shen-Orr.etal:2002}. If such a function
is required repeatedly in different parts of the network, there is
selection pressure for the creation and maintenance of similar
topologies in different parts of a network. Such 
 \textit{network motifs}~\cite{Milo.etal:2002,Shen-Orr.etal:2002}
are families of subgraphs
distinguished from the null model by mutual correlations between subgraphs, 
see Fig. \ref{fig_stats} b).

To quantify these correlations, we need to specify the parts of the
network with correlated patterns. We define a {\em graph alignment}
$\A$ by a set of several node subsets $\Areg^\alpha$ ($\alpha = 1, \dots,
p$), each containing the same number of $n$ nodes, and a specific
order of the nodes $\{r_1^\alpha, \dots, r_n^\alpha\}$ in each node subset.
An alignment associates each node in a node subset with exactly one node in
each of the other node subsets.  The alignment can be visualized by $n$ ``strings'',
each connecting $p$ nodes as shown in Fig.~\ref{fig_stats}(b).

An alignment specifies a pattern $\ahat^\alpha \equiv \ahat
(\Areg^{\alpha},\A)$ in each node subset. For any two aligned subsets of nodes, 
$\Areg^\alpha$ and $\Areg^\beta$, we can define the {\em pairwise mismatch} 
of their patterns
\begin{equation}
\label{pairwise_mismatch}
M(\ahat^\alpha, \ahat^\beta) =
  \sum_{i,i'=1}^n [\hat{a}_{ii'}^\alpha (1 - \hat{a}_{ii'}^\beta) +
            (1 - \hat{a}_{ii'}^\alpha) \hat{a}_{ii'}^\beta ] \ .
\end{equation}
The mismatch is a Hamming distance for aligned patterns. The average
mismatch over all pairs of aligned patterns is termed the 
{\em fuzziness} of the alignment.

Frequently network 
motifs also have an enhanced number of 
internal links~\cite{Shen-Orr.etal:2002,Milo.etal:2002}, 
providing the possibility of feedback or 
other faculties not available to tree-like patterns. 
An ensemble describing $p$ node subsets with correlated patterns  
$\ahat^1, \dots, \ahat^p$ with an enhanced number of links is 
given by
\begin{eqnarray}
\label{Qmotifs}
\lefteqn{Q_{\mu,\sigma} (\ahat^1, \dots, \ahat^p) 
 = Z_{\mu,\sigma}^{-1}\prod_{\alpha = 1}^p P_0 (\ahat^{\alpha})} \\
&& \times
\exp \left [ - \frac{\mu}{2p} \sum_{\alpha,\beta=1}^p
M(\ahat^{\alpha},\ahat^{\beta}) 
+ \sigma \sum_{\alpha=1}^p L( \ahat^{\alpha} )
\right ]
 \ .  \nonumber
\end{eqnarray}

The parameter $\mu \geq 0$ biases the ensemble~(\ref{Qmotifs})  
towards patterns with small mutual mismatches 
$M(\ahat^{\alpha},\ahat^{\beta})$. 

Given the null model (\ref{sigma_enhanced}) and the model 
(\ref{Qmotifs}) with correlated patterns, we obtain 
a log-likelihood score for network motifs 
\begin{eqnarray}
\label{scoremotifs}
\lefteqn{S(\A,\mu,\sigma)}
\nonumber \\
&  = & 
\log  \left (\frac{Q_{\mu,\sigma} (\ahat^1, \dots, \ahat^p)}{
            P_0 (\ahat^1, \dots, \ahat^p)} \right)
\nonumber \\        
& = & - \frac{\mu}{2p} \sum_{\alpha,\beta=1}^p 
M(\ahat^{\alpha},\ahat^{\beta})
+\sigma \sum_{\alpha=1}^p L( \ahat^{\alpha})
 \nonumber \\
 & & -\log Z_{\mu,\sigma} \ .
\end{eqnarray}
High-scoring alignments $\A$ indicate \textit{bona fide} network motifs. The first
and second term reward alignments with a small mutual mismatch and a high number of internal
links, respectively. The term
$\log Z_{\sigma,\mu}$ acts as a threshold assigning a
negative score to alignments with too high fuzziness or too few 
internal links.

Again, both the alignment
$\A$ and the scoring parameters $\mu$ and $\sigma$ are 
\textit{a priori} undetermined. For given scoring parameters, the maximum-score alignment 
\be
\A^\star(\mu,\sigma)= \argmax_{\A}S(\A,\mu,\sigma) 
\ee 
occurs at some finite value of the number of subgraphs
$p^\star(\mu,\sigma)$. 

The scoring parameters $\mu$ and $\sigma$ can again be determined by
maximum likelihood, which corresponds to maximizing the score
$S(\A^{\star}(\mu,\sigma),\mu,\sigma)$ with respect to the scoring
parameters. By differentiating (\ref{scoremotifs})
with respect to the scoring parameters one finds that at
$\mu=\mu^{\star}$ and $\sigma=\sigma^{\star}$ 
the model~(\ref{Qmotifs}) fits the maximum-score network motifs: 
the expectation values of the internal number of links and the
fuzziness equal the corresponding values of the maximum-score alignment. 
 
\para{Network motifs in regulatory networks} We now apply  
the scoring function (\ref{scoremotifs}) to the identification of network 
motifs in the gene regulatory network of 
{\em E. coli}, taken from~\cite{Shen-Orr.etal:2002}. A full account and a score-maximization 
algorithm are given in \cite{BergLaessig:2004}.

\begin{figure}[tbh!]
(a)
\includegraphics*[width=.8 \linewidth]{pscan.eps}


(b)
\includegraphics*[width=.96\linewidth]{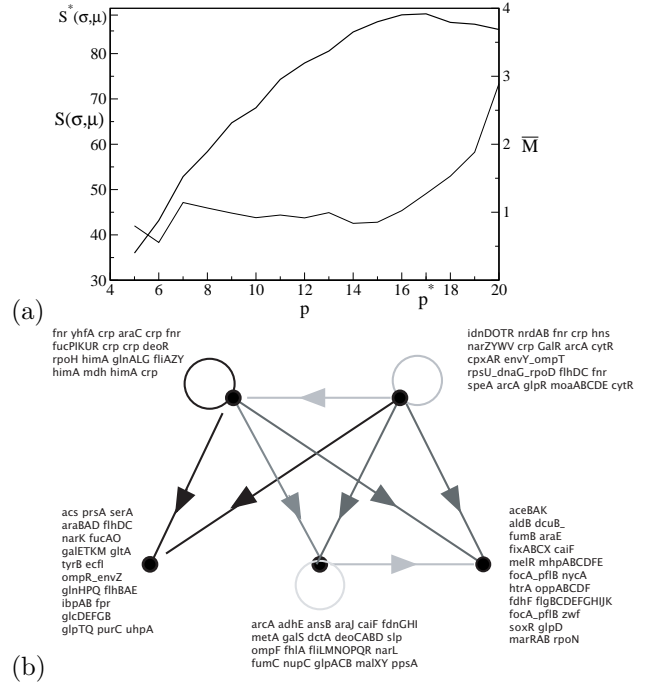}
\caption{ \label{figopt} \small
{\bf Motifs in the regulatory network of \textit{E. coli}}.
(a) Score optimization at fixed scoring parameters $\sigma = 3.8$ and 
$\mu = 4.0$ for subgraphs
of size $n=5$. The total score $S$ (thick line) and the fuzziness 
${\overline M}$ (thin line) are shown for the highest-scoring 
alignment of $p$ subgraphs, plotted as a function of $p$. 
(b)
The consensus motif of the optimal alignment, and the 
identities of the genes involved. The alignment 
consists of $18$ subgraphs sharing at most one node. 
The $5$ grey values correspond to the consensus motif $\overline{a}$ in the
range $0.1-0.2$, $0.2-0.4$, $0.4-0.6$, $0.6-0.8$, $0.8-0.9$.  
}
\end{figure}

We first investigate the properties of the maximal score alignment at
fixed scoring parameters.  Fig.~\ref{figopt}~(a) shows the score $S$
and the fuzziness ${\overline M}$ for the highest-scoring alignment
with a prescribed number $p$ of subgraphs, plotted against $p$.  The
fuzziness increases with $p$, and the score reaches its maximum
$S^{\star}(\sigma, \mu)$ at some value $p^{\star}(\sigma, \mu)$. For $p < p^{\star}
(\sigma, \mu)$ the score is lower, since the alignment contains fewer
subgraphs and for $p > p^{\star} (\sigma, \mu)$ it is lower since the
subgraphs have higher mutual mismatches.

The optimal scoring parameters $\mu$ and $\sigma$ are again  
inferred by maximum likelihood. The resulting optimal 
alignment $\A^\star \equiv \A^\star(\mu^\star,\sigma^\star)$ 
is shown in Fig.~\ref{figopt}~(b) using the 
so-called \textit{consensus motif}
\be
\label{cbar}
\overline {\bf a} = \frac{1}{p} \sum_{\alpha =1}^p {\ahat}^\alpha(\A^\star) \ .
\ee 
The consensus motif is a {\em probabilistic pattern}; the entry $\overline{a}$ 
denotes the probability that a given binary link is present in the aligned
subgraphs. The motif shown in Fig.~\ref{figopt}~(b) consists of 
$2+3$ nodes forming an input and an output layer, with links 
largely going from the input to the output layer. Most 
genes in the input layer code for transcription factors or 
are involved in signaling pathways. The output layer mainly consists 
of genes coding for enzymes.  


\subsection*{5. Cross-species analysis of networks}

The motifs discussed above show correlation without sharing a 
common evolutionary history. Larger functional units may be
distinguished by
their evolutionary conservation. Thus, we expect 
parts of the network to maintain their topology and to form a conserved 
core, while other parts show a more rapid turnover of both 
nodes and interactions, see Figure~\ref{fig_stats}c). This
conservation can be detected as topological correlation
across species. 

We assume that organisms evolve independently after
speciation, leading to divergence in their network links
as well as in the overall similarity of the nucleotide
sequences, the structure of proteins, and
the biochemical role of a metabolite. The relationship between
link and node similarity is non-trivial: genes may retain their function 
and their interactions with other genes despite considerable 
sequence divergence. On the other hand, the change of a few nucleotides 
can create or destroy a binding site, implying that genes with 
high overall sequence similarity may have entirely different interactions. 
Hence, the cross-species analysis has to take into account information 
from both links and nodes. 

A log-likelihood score assessing the 
link statistics of node subsets in network $A$ and 
in network $B$ follows 
directly from~(\ref{scoremotifs}). This \textit{link score} is given by 
\begin{eqnarray}
\label{scorecross_binary}
\lefteqn{S^\ell(\A,\mu,\sigma_A,\sigma_B)=
- \mu M(\ahat,\bhat)}
 \\
&&+ \sigma  \left( L(\ahat)+ L(\bhat) \right)
 -\log Z_{\mu,\sigma} \nonumber \ . 
\end{eqnarray}

To assess the similarity of nodes, we consider 
a measure $\theta_{ij}$, which describes the similarity of 
node $i$ in network $A$ and node $j$ in network
$B$. The node similarity measure may be a percentage sequence identity, or a distance
measure of protein structures. The information on node similarity can
be incorporated into the alignment score by contrasting a null model 
with a model describing a statistics where node similarity is correlated with the
alignment. To construct the null model, we assume that node
similarities $\theta_{ij}$ for different node pairs $i,j$ are
identically and independently distributed and denote their
distribution by $p_0^n (\theta_{ij})$.
The model describing cross-species correlations has to take into 
account that the distribution of node
similarities between aligned pairs of nodes follows a different
statistics (typically generating higher values of $\theta$), denoted
by $q^n_1(\theta)$. The distribution of pairwise similarity
coefficients between one aligned node and nodes other than its
alignment partner is denoted by $q^n_2 (\theta)$.

\begin{figure}[t!]
\includegraphics*[width=1 \linewidth]{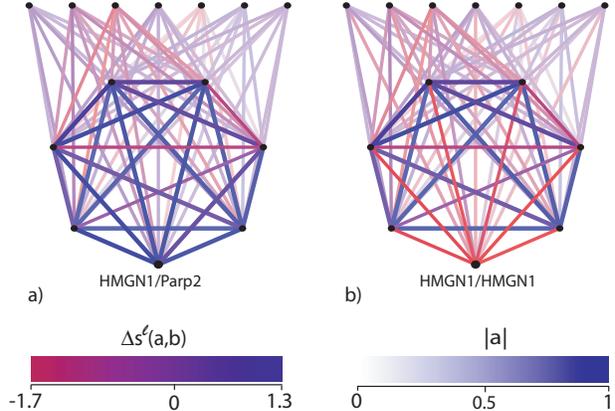}
\caption{\label{fig_netpics} \small
{\bf Cross-species network alignment shows conservation of gene clusters.} 
(a) 
$7$ genes from a cluster of co-expressed genes (circle) together with
$7$ random genes outside the cluster (straight line). Each node
re\-pre\-sents a pair of aligned genes in human and mouse. The intensity
of a link encodes the correlation value $a$ in human. The color
indicates the evolutionary conservation of a link, with blue hues
indicating strong conservation. Th conservation is quantified by the
excess link score contribution, $\Delta s^\ell$, defined as the link
score minus the average link score of links with the same correlation
value.
(b) 
The same cluster, but with human-HMGN1 ``falsely'' aligned to its ortholog
mouse-HMGN1, with the red links showing the poor expression overlap of 
this pair of genes. 
}
\end{figure}

Assuming that the statistics of links and
nodes similarities are uncorrelated for a given alignment, a simple calculation analogous to
(\ref{bayes_score}) yields the log-likelihood score
 \be
S(\A) = S^\ell (\A) + S^n (\A) \ ,
\label{Stot}
\ee
with the information from node similarity contributing a \textit{node score}
\be
\label{Sn_general}
S^n (\A) = 
\sum_{i\in \Areg} s^n_1 (\theta_{ii}) + 
\sum_{ 
\scalebox{0.75}{
 \parbox{2.5cm}
{$i \in \Areg, j \neq i$ \\
 $j \in \Breg , i \notin \Areg$ }
} } 
s^n_2 (\theta_{ij}) 
\ee
and $s^n_1 (\theta) \equiv \log \left( q^n_1(\theta)/p_0^n (\theta)
\right)$ and $s^n_2 (\theta) \equiv \log \left( q^n_2(\theta)/p_0^n
  (\theta) \right)$.

The scoring parameters entering (\ref{Stot}) need to be
determined from the data. Provided there are not too many scoring
parameters, this can again be done by maximum likelihood as outlined
in the preceding sections. Particular examples are networks with
binary links and course-grained measures of sequence similarity. (As
an extreme case, node similarity may be considered a binary variable,
when nodes either have significant similarity or not. Then the
ensembles describing the node statistics are each described by a
single variable, see~\cite{BergLaessig:06} for details.)

\para{Alignment of co-expression networks}
We compare co-expression networks of \textit{H. sapiens} and \textit{M. musculus}. In
co-expression networks, the weighted link $a_{ii'} \in [-1,1]$ between a
pair of genes $i,j$ is given by the correlation coefficient of their gene 
expression profiles measured on a microarray chip.  Genes which
tend to be expressed under similar conditions thus have positive
links. The score~(\ref{scorecross_binary}) can easily be generalized 
to weighted interactions, see \cite{BergLaessig:06}. 

The data of Su et al.~\cite{Su.etal:2004} was used to construct
networks of $\sim 2000$ housekeeping genes. Human-mouse orthologs were
taken from the Ensembl database~\cite{Ensembl:2005}.  Details on the
algorithm to maximize the score (\ref{scorecross_binary}) are given
in~\cite{BergLaessig:06}.

We focus on strongly conserved parts of the two networks. 
Figure~\ref{fig_netpics} shows a cluster of co-expressed genes which 
is highly conserved between human and mouse (link conservation is shown 
in blue, changes between the links in red). 

With one exception, the aligned gene pairs in this cluster have
significant sequence similarity and are thought to be orthologs,
stemming from a common ancestral gene. The exception is the aligned
gene pair human-HMGN1/mouse-Parp2. These genes are aligned due to
their matching links, quantified by a high contribution to the link
score~(\ref{scorecross_binary}) of $S^\ell = 25.1$. The ``false''
alignment human-HMGN1/mouse-HMGN1 respects sequence similarity but
produces a link mismatch ($S^\ell = -12.4$); see 
Fig.~\ref{fig_netpics}(b). Human-HMGN1 is known to be involved in
chromatin modulation and acts as a transcription factor. The network
alignment predicts a similar role of Parp2 in mouse, which is distinct
from its known function in the poly(ADP-ribosyl)ation of nuclear
proteins. The prediction is compatible with experiments on the effect of
Parp-inhibition, which suggest that Parp genes in mouse play a role in
the chromatin modification during development~\cite{Imamura:2004}.

\subsection*{6. Towards an evolutionary theory}

Different parts of biological networks have different
functions. Here we have
applied a statistical approach to the detection of network clusters,
network motifs, and cross-species correlations.  But the detection of
deviations from a global background statistics has a wider
perspective, which includes
the connection between different type of networks, the link
between network topology and the underlying sequence, and
spatiotemporal changes of biological networks. 
From an evolutionary point of view, these deviations are
created and maintained by selection pressures which are both
non-homogeneous and correlated across the network. A
quantitative theory of biological networks will thus require a
synthesis of network statistics and population genetics, a
largely outstanding task to date. Here we give a brief
outlook on some of the challenges ahead. 

\para{Genetic interactions between different links} Biological
function is typically tied to modules consisting of several nodes and 
links.  As a result, there are correlations between links across
different species: a species with a certain function will tend to have
all links associated with the specific function, a species lacking the 
function will tend to have none of the corresponding links.  The
network motifs discussed above are only a special case of this
phenomenon. With data on biological networks becoming available for an
increasing number of species, it will become feasible to infer these
correlations and the corresponding functional modules from data.
Scoring functions constructed to detect genetic interactions in
multiple alignments will play an important role in this undertaking.

\para{Gene duplications} Following the duplication of a gene, the 
daughter genes have the same function and same interactions with other
genes. Independent evolution of the two genes may lead to the
non-functionalisation, and even the loss of one of the duplicates, or to
subfunctionalisation, with different functional roles being divided
among the two copies~\cite{LynchO'HelyWalshForce:2001}. Tracing the dynamics of 
gene duplication at the level of
interaction networks
gives insights into the evolutionary dynamics of
networks~\cite{BergLaessigWagner:2004,Chung.etal:2006}. Scoring for jointly conserved
subgroups of links can be used to identify the different functional
modules a gene is involved in.  This can be done both at the level of
single species, as well as in a cross-species analysis, where
gene duplications introduce one-to-many and many-to-many
alignments.

\para{Neutral and selective dynamics} Biological networks show a great deal of
plasticity, since the same biological function can be carried out by
different networks (see e.g.\ \cite{Tanayetal:2005}).  This flexibility leads to
neutral evolution as a population explores the space of networks
corresponding to a given function.  On the other hand, networks may
change as a new functionality is acquired, or because of changing
environmental conditions.  Disentangling neutral moves and changes
under selection is possible by contrasting inter-species
variability with intra-species
variability~\cite{McDonaldKreitman:1991}. Inferring the modes of
network evolution and the relative weights of neutral and selective
dynamics remains an outstanding challenge for experiment and theory.

\noindent
{\bf Acknowledgments:}
This work was supported through DFG grants SFB/TR 12, SFB 680, and 
BE 2478/2-1. We thank David Arnosti, Daniel Barker, Leonid Mirny, 
and Nina White for discussions.

\subsection*{Appendix: \\
Bayesian analysis of network data}
The detection of deviations from a null model can be formulated as a 
problem of deciding between alternative hypotheses. The first
hypothesis is that a given node subset follows the statistics of the null model.
The alternative hypothesis is that the node subset follows a statistics
different from the null model. This statistics is called the 
$Q$-model. 

The choice between these two alternatives can be formulated
probabilistically, by considering the \textit{posterior probability}
$P(Q|\ahat,\A)$.  It describes the probability that the node subset(s)
specified by $\A$ follow the $Q$-model (hypothesis $Q$), rather than
the null model (null-hypothesis $P_0$). Denoting any prior
knowledge we may have about the probability with which the two 
alternatives occur by $P(Q)$ and $P(P_0)$, respectively, one may use
Bayes' theorem to find 
\bea P(Q |\ahat,\A)
&=& \frac{P(\ahat|Q,\A) P(Q)} {P(\ahat|\A)} \\
&=& \frac{P(\ahat|Q,\A) P(Q)}
{P(\ahat|P_0,\A)P(P_0)+P(\ahat|Q,\A)P(Q)}
\nonumber \\
&=&\frac{e^{S'(\A)}} {1+e^{S'(\A)} } \nonumber \ .  
\eea
$P(\ahat|Q,\A)$ gives the probability of generating patterns $\ahat$
under the $Q$-model (given, for instance, by ~(\ref{Qcluster}) or
by~(\ref{Qmotifs})). $P(\ahat|P_0,\A)$ gives the probability of
generating the same pattern under the
null model~(\ref{sigma_enhanced}).  The posterior probability is thus
a monotonously increasing function of the {\em
  log-likelihood score} given by 
\bea
\label{bayes_score2}
S'(\A)&=&\log \left( \frac{P(\ahat|Q,\A)}{P(\ahat|P_0,\A)} \right) + 
\log \left( \frac{P(Q)}{P(P_0)} \right)\ \nonumber \\
&=&S(\A) + {\mbox const.}
\eea 
Hence the score $S(\A)$ defined in~(\ref{bayes_score}) has a sound
theoretical foundation: it is a measure of the
posterior probability that the node subset specified by $\A$
follows the $Q$-model rather than the null model.

This simple picture needs to be extended when the parameters $m$ of
the $Q$-model and the alignment $\A$ are unknown and are considered
``hidden'' variables to be determined from the data. We construct a
model for the entire network with adjacency matrix $\a$, with pattern
$\ahat(\A)$ following the $Q$-model, the remainder of the network
following the null model
\be
P(\a| \A,m)= Q(\ahat|\A,m) P_0({\bf \tilde a}|\A) 
\ .  
\ee 
The matrix of links
between nodes which are not both part of $\A$ is denoted by ${\bf
  \tilde a}$.  Using Bayes' theorem one can write the posterior
probability of $\A$ and $m$, i.e.\ the conditional probability of the
hidden variables, in the form
\be P(\A,m | \a) =
\frac{Q(\a| \A,m) P (\A, m)}{
   \sum_{\A,m} Q(\a | \A,m)P (\A, m) } \ .
\label{full_posterior}
\ee 
We assume the prior probability $P (\A,m)$ to be flat. Dropping
the terms independent of $\A$ and $m$, the optimal alignment
$\A^\star$ is obtained by maximizing the posterior probability $Q(\A |
\a) \sim \sum_m Q(\a| \A,m)$ with respect to $\A$ and similarly the
optimal scoring parameters $m^\star$ by maximizing $Q(m | \a) \sim
\sum_\A Q(\a| \A,m)$ with respect to $m$. In the so-called Viterbi
approximation, $\A^\star$ and $m^\star$ are inferred by jointly
maximizing $Q(\a,\b,\bth | \A,m)$ with respect to $\A$ and
$m$. Assuming the sum $\sum_{\A,m} Q(\a | \A,m)$ can be split into the
term stemming from $\A^\star,m^\star$ and a remainder $\sum_{\A\neq
  \A^\star,m\neq m^\star} Q(\a | \A,m)\sim P_0(\a)$, the
posterior probability~(\ref{full_posterior}) can again be written in
the form of~(\ref{bayes_score2}). In this approximation, the
maximum-score alignment and the optimal scoring parameters are
determined by the maximum of the log-likelihood score
~(\ref{bayes_score}) over the alignments and over the scoring
parameters.


\end{document}